\documentclass[12pt,english,floatfix,superscriptaddress,aps,prd,preprint,showpacs]{revtex4}
\usepackage{amsmath}
\usepackage{amssymb}
\usepackage{amsbsy}
\usepackage{amsfonts}
\usepackage{amsopn}
\usepackage{amstext}
\usepackage{esint}
\usepackage{amssymb}
\usepackage{amsfonts}
\usepackage{amsmath}
\usepackage[english]{babel}
\usepackage{color}
\usepackage[dvips]{epsfig}
\usepackage[dvips]{graphicx}
\usepackage{float}
\usepackage{units}
\usepackage{textcomp}
\usepackage{babel}
\usepackage[dvips]{graphicx}
\usepackage{subfig}
\usepackage{hyperref}
\usepackage{slashed}

\newcommand{\incps}[5]{\includegraphics[#2,#3][#4,#5]{#1}}

\begin{document}

\title{Radiative corrections in bumblebee electrodynamics}

\author{R. V. Maluf}
\email{r.v.maluf@fisica.ufc.br}
\affiliation{Universidade Federal do Cear\'a (UFC), Departamento de F\'isica, Campus do Pici, Fortaleza - CE, C.P. 6030, 60455-760 - Brazil}


\author{J. E. G. Silva}
\email{jgsilva@indiana.edu}
\affiliation{Indiana University Center for Spacetime Symmetries, Bloomington,
Indiana 47405, USA}


\author{C. A. S. Almeida}
\email{carlos@fisica.ufc.br}
\affiliation{Universidade Federal do Cear\'a (UFC), Departamento de F\'isica, Campus do Pici, Fortaleza - CE, C.P. 6030, 60455-760 - Brazil}


\begin{abstract}
We investigate some quantum features of the bumblebee electrodynamics in flat spacetimes. The bumblebee field is a vector field that leads to a spontaneous Lorentz symmetry breaking. For a smooth quadratic potential, the massless excitation (Nambu-Goldstone boson) can be identified as the photon, transversal to the vacuum expectation value of the bumblebee field. Besides, there is a massive excitation associated with the longitudinal mode and whose presence leads to instability in the spectrum of the theory. By using the principal-value prescription, we show  that no one-loop radiative corrections to the mass term is generated. Moreover, the bumblebee self-energy is not transverse, showing that the propagation of the longitudinal mode can not be excluded from the effective theory.
\end{abstract}


\maketitle

\section{Introduction}

At the Planck scale, several theories consider the possibility of the quantum spacetime structure leads to violation of the Lorentz symmetry. In the noncommutative theories, the spacetime has a minimum length \cite{noncommutativegeometry} whereas in the Ho\v{r}ava-Lifshitz  gravity the space and time covariance is no longer valid \cite{horava}. Furthermore, the additional vector and tensor fields in the string theory may acquire a non-vanishing vacuum expectation value yielding to a preferred direction in the spacetime \cite{KS,espontaneous lorentz violation string}.

An effective quantum field theory that accounts for the Lorentz violating effects and preserves the gauge structure of the standard model (SM) of the fundamental interactions is called the standard model extension (SME) \cite{sme}. The SME neatly incorporates violation of Lorentz symmetry by adding terms to the standard model Lagrangian, which explicitly breaks the Lorentz symmetry at the particle frame. The Lorentz violating terms are constructed from the vacuum expectation value of the tensor fields that are constant background fields. The search for Lorentz violating signals covers all interactions sectors; the gauge sector \cite{gauge,gauge2,gauge3,gauge4}, the fermion sector \cite{fermion,fermion2}, and extensions involving gravity \cite{KosteleckyG1,Bluhm2005,BluhmFung,Petrov,Pereira2011,Boldo,MalufGravity1,MalufGravity2,PetrovGodel}. For a comprehensible analysis of the Lorentz violating data, see for instance the Ref. \cite{Kostelecky:2008ts}.

A dynamical violation of the Lorentz symmetry can be achieved by means of spontaneous symmetry breaking mechanism \cite{KS,KosteleckyG1,Bluhm2005,BluhmFung,
Seifert,altschul,Bluhm:2008yt,Hernaski:2014jsa}. The simplest field theories involving a vector that acquires nonzero vacuum expectation values are the so-called bumblebee models \cite{KS,KosteleckyG1,Bluhm2005,BluhmFung}. The Lorentz violating is trigged by a vector field, called the bumblebee field, whose minimum of the potential gives rise to the background field. Amongst the possible choices for the potential are the smooth quadratic \cite{KS,Bluhm2005,BluhmFung}, the Lagrange-multiplier \cite{BluhmFung} and the nonpolynomial potentials \cite{altschul}.  Kosteleck\'{y} and Samuel proposed the usual Maxwell-like kinetic term for the bumblebee field and a smooth quadratic potential \cite{KS}. The presence of the potential also breaks the gauge symmetry of the vector field.

In the Kosteleck\'{y}-Samuel (KS) model, the quadratic bumblebee Lagrangian expanded around the vacuum value $b_{\mu}$ has the form of the Maxwell Lagrangian added with an axial gauge fixing term \cite{KS,Bluhm2005,BluhmFung,altschul,Bluhm:2008yt}. Furthermore, the excitations of the bumblebee field can be cast into two classes: two Nambu-Goldstone (NG) modes, transverse to the vacuum expected value of the bumblebee field, and one massive or longitudinal mode \cite{KS,Bluhm2005,BluhmFung,altschul,Bluhm:2008yt}. At tree level in the linear regime, the two NG modes can be identified with the two polarization modes of the photon \cite{BluhmFung}. In turn, the massive mode appears as a tachyonic ghost excitation and the associated Hamiltonian of the model is unbounded below \cite{Bluhm:2008yt,Carroll2009}. Nevertheless, a appropriated choice of the initial conditions for the field configurations can yield regions of phase space that are ghost-free and have Hamiltonian positive \cite{Bluhm:2008yt}. 

An important question is how the quantum effects affect the stability of the KS theory and the propagation of the unphysical modes. In the work of Ref. \cite{Hernaski:2014jsa}, the canonical quantization of the KS model in flat Minkowski spacetime was addressed. The St\"{u}ckelberg method was employed to define an extended Fock space such that the second-class constraints are converted into the first-class ones. In the restricted Fock space for the physical states, the free KS model turns out to be unitary and equivalent to the Maxwell electrodynamics in the temporal gauge \cite{Hernaski:2014jsa}. However, to the best of our knowledge, the radiative corrections induced by the interaction terms in the KS model were not addressed in the literature yet.

In this letter, we study the radiative corrections to the bumblebee field in the KS model in the flat spacetime. We started with the analysis of the free propagation modes of the bumblebee field at tree level. We show that the massive longitudinal excitation does not represent a physical propagating mode. In the sequel, the quantum effects are accessed exploring the similarity between the bumblebee model at the linear approximation and the Maxwell theory in the axial gauge. We evaluate the one-loop radiative corrections due to the bumblebee self-energy and we find that no correction for the mass term of the longitudinal mode is generated. However, the transversality condition $p_\mu\Pi^{\mu\nu}(p)=0$ is not satisfied,  which confirms that the massive mode is naturally going to be excited by the interactions terms.

This letter is organized as the following. In Sec. \ref{sec:theoretical-model}, we define and review the main properties of the KS model and study the free propagation of the bumblebee field in the absence of matter. In Sec. \ref{radiativecorrections}, we study the one-loop radiative corrections of the bumblebee self-energy. Finally, our conclusions, as well as perspectives are outlined in section \ref{sec:Conclusions}.

\section{Bumblebee electrodynamics \label{sec:theoretical-model}}

We begin presenting some classical results on a particular class of bumblebee models, namely the
Kosteleck\'{y}-Samuel (KS) model.

The Lagrangian for the KS model that describes the dynamics of the bumblebee field is given by \cite{KS,KosteleckyG1,Bluhm2005,BluhmFung}
\begin{equation}
\label{flatlagrangian}
\mathcal{L}_{B}=-\frac{1}{4}B_{\mu\nu}B^{\mu\nu}-\frac{\lambda}{4}(B^{\mu}B_{\mu}\pm b^{2})^{2}-B^{\mu}J_{\mu},
\end{equation}where $\lambda$ is a dimensionless positive coupling, $b^{2}$ is
a positive constant with squared mass dimension, $J_{\mu}$ is supposed
to be a conserved current formed of matter fields that are also the
source for the $B_{\mu}$ field, and the field-strength tensor $B_{\mu\nu}$
is defined as
\begin{equation}
B_{\mu\nu}=\partial_{\mu}B_{\nu}-\partial_{\nu}B_{\mu}.
\end{equation}

The smooth quadratic potential term $V=-\frac{\lambda}{4}(B^{\mu}B_{\mu}\pm b^{2})^{2}$
is responsible for triggering the mechanism of spontaneous Lorentz
violations. The bumblebee field takes on a nonzero vacuum
value $\left\langle B_{\mu}\right\rangle =b_{\mu}$ for a local minimum
at $B^{\mu}B_{\mu}\pm b^{2}=0$, such that $b^{\mu}b_{\mu}=\mp b^{2}$ with the $\mp$ sign meaning if $b^{\mu}$ is spacelike or timelike \cite{KS,KosteleckyG1,Bluhm2005}. Also, note that the potential ensures explicit violation of $U(1)$ gauge symmetry.

As discussed in Ref. \cite{BluhmFung}, in theories with spontaneous Lorentz violation, the potential propagating modes can be classified into five types: gauge modes, Nambu-Goldstone (NG), massive
modes, Lagrange-multiplier modes, or spectator modes. In particular, for the theory defined by the Lagrangian density in Eq. \eqref{flatlagrangian}, only the NG and massive modes are present. The NG modes arise if the excitations satisfy the condition $V'(X)=0$, where the prime denotes the derivative with respect to $X=B^{\mu}B_{\mu}\pm b^{2}$. Consequently, the massive mode is an excitation associated with a non-minimal value of the potential, and it is observed when $V'(X)\neq0$.

The stability and the unitarity of the Hamiltonian associated with the Lagrangian \eqref{flatlagrangian} were studied in Refs. \citep{Bluhm:2008yt,Hernaski:2014jsa,Carroll2009}. Besides the massless NG mode, the KS model also possesses a propagating massive tachyonic excitation which leads to instabilities. However, one can consistently choose a set of constraints which reduce the phase space to a region where the Hamiltonian is positively definite \citep{Bluhm:2008yt}. In this restricted phase space the tachyon does not propagate and the free model is classically equivalent to the Maxwell theory in a nonlinear gauge \citep{Bluhm:2008yt}. By means of the St\"{u}ckelberg method and a suitable choice of the creation and annihilation operators, it is possible to define a reduced Fock space which is ghost-free and whose physical states have positive Hamiltonian \citep{Hernaski:2014jsa}.

In the sequel, we use the Lagrangian approach to show that no physical propagating massive mode can be achieved.

Since our main objective is to study the dynamics of the bumblebee field
$B_{\mu}$ around the vacuum, we adopt the following decomposition
\begin{equation}
B_{\mu}=b_{\mu}+\beta_{\mu}.\label{eq:B-linearized}
\end{equation}

In terms of the excitation $\beta_{\mu}$, the bumblebee Lagrangian \eqref{flatlagrangian} can be rewritten as
\begin{equation}
\tilde{\mathcal{L}}_{B}=-\frac{1}{4}\beta_{\mu\nu}\beta^{\mu\nu}-\frac{\lambda}{4}\left(4\beta_{\mu}b^{\mu}\beta_{\nu}b^{\nu}+\beta_{\mu}\beta^{\mu}\beta_{\nu}\beta^{\nu}+4\beta_{\mu}\beta^{\mu}\beta_{\nu}b^{\nu}\right)-\beta_{\mu}J^{\mu}-b_{\mu}J^{\mu},
\label{eq:Lagra1}
\end{equation}
where $\beta_{\mu\nu}=\partial_{\mu}\beta_{\nu}-\partial_{\nu}\beta_{\mu}$ is a field strength for the bumblebee excitation \cite{Hernaski:2014jsa}.
The interaction vertices of the theory are generated by the trilinear
and quadrilinear terms of \eqref{eq:Lagra1}. Note that a mass term arise naturally and involves the mass matrix $m_{\mu\nu}=2\lambda b_{\mu}b_{\nu}$. From the quadratic terms of the  Lagrangian density \eqref{eq:Lagra1}, we can extract the free  bumblebee propagator given by
\begin{equation}
\label{axialgaugepropagator}
D_{F}^{\mu\nu}(p,b)=-\frac{i}{p^{2}+i\epsilon}\left[g^{\mu\nu}-\frac{(p^{\mu}b^{\nu}+p^{\nu}b^{\mu})}{b\cdot p}+\frac{(p^{2}+2\lambda b^{2})}{2\lambda(b\cdot p)^{2}}p^{\mu}p^{\nu}\right],
\end{equation} which is similar to the gauge field propagator in the axial-gauge \cite{altschul,Hernaski:2014jsa,pv,pv2,Leibbrandt:1987qv,LeibbrandtBook}.

The first term in \eqref{axialgaugepropagator} has a pole at $p^2=0$, and it represents a massless excitation. Since the transverse mode is massless, we identify this term with the transverse mode. The double pole $b\cdot p=0$ indicates a non-physical mode induced by the Lorentz-violating term and whence, it is naturally associated with the massive longitudinal excitation.

In order to analyse the free propagation of the bumblebee field, we focus our
attention only on the quadratic terms of $\tilde{\mathcal{L}}_{B}$. The equation of motion reduces to 
\begin{equation}
\partial^{\mu}\beta_{\mu\nu}-2\lambda\beta^{\mu}b_{\mu}b_{\nu}=0.\label{eq:mov1}
\end{equation}

Since the background vector $b_{\mu}$ defines a preferred direction in space, we can split the excitations $\beta_{\mu}$ into transverse $(A_{\mu})$ and longitudinal $(\beta)$ modes by means of the orthogonal projection operators \cite{,Bluhm2005,BluhmFung}
\begin{equation}
P_{\mu\nu}^{||}=\frac{b_{\mu}b_{\nu}}{b^{\alpha}b_{\alpha}}\ \ \ \ \ \ \mbox{and}\ \ \ \ \ \ P_{\mu\nu}^{\perp}=g_{\mu\nu}-\frac{b_{\mu}b_{\nu}}{b^{\alpha}b_{\alpha}},\label{eq:project}
\end{equation}
such that
\begin{eqnarray}
\beta_{\mu} & \equiv & A_{\mu}+\beta\hat{b}_{\mu},\label{modeTL}\\
 A_{\mu} & = & P_{\mu\nu}^{\perp}\beta^{\nu}\ \ \mbox{(transverse mode)},\label{TransverseMode}\\
\beta\hat{b}_{\mu}  & = & P_{\mu\nu}^{||}\beta^{\nu}\ \ \mbox{(longitudinal mode),}
\end{eqnarray}
with $A_{\mu}b^{\mu}=0$ and $\hat{b}_{\mu}=b_{\mu}/\sqrt{b^{2}}$,
so that $\hat{b}^{\mu}\hat{b}_{\mu}=\mp1$. With these definitions the equation of motion \eqref{eq:mov1} can be written as
\begin{equation}
\partial^{\mu}F_{\mu\nu}+\square\beta\hat{b}_{\nu}-\partial_{\nu}\partial^{\mu}\beta\hat{b}_{\mu}-2\lambda\beta\hat{b}^{\mu}b_{\mu}b_{\nu}  = 0,\label{eq:mov2}
\end{equation}
where $F_{\mu\nu}=\partial_{\mu}A_{\nu}-\partial_{\nu}A_{\mu}$. Applying
$\partial^{\nu}$ on Eq. \eqref{eq:mov2} we obtain the following
constraint 
\begin{equation}
b_{\nu}\partial^{\nu}\beta=0.\label{eq:constraint1}
\end{equation}

Substituting the constraint \eqref{eq:constraint1} in \eqref{eq:mov2}
and using the projectors \eqref{eq:project} to separate the equations
of $A_{\mu}$ and $\beta$, we come to the following equations of motion for each mode: \begin{eqnarray}
\square A_{\mu}-\partial_{\mu}\partial^{\nu}A_{\nu}+\frac{1}{b^{2}}b_{\mu}b^{\nu}\partial_{\nu}\partial^{\lambda}A_{\lambda} & = & 0,\\
\square\beta-2\lambda b^{\alpha}b_{\alpha}\beta-\frac{1}{b^{\alpha}\hat{b}_{\alpha}}b^{\mu}\partial_{\mu}\partial^{\nu}A_{\nu} & = & 0.
\end{eqnarray}

The constraint $b_{\nu}\partial^{\nu}\beta=0$ imposes the additional condition for the massive mode $\beta$ in the momenta space
\begin{equation}
\label{betaconstrain}
b^{\mu}p_{\mu}=0.
\end{equation}
The condition \eqref{betaconstrain} along with the equation of motion for the $\beta$ yield to the following dispersion relation for the massive mode
\begin{equation}
p^{2}+2\lambda b^{\alpha}b_{\alpha}=0,
\end{equation}
such that the associated mass to this excitation is give by
\begin{equation}
\label{massivemodemass}
M_{\beta}^{2}=-2\lambda b^{\alpha}b_{\alpha}.
\end{equation}

The constraint $(\ref{betaconstrain})$ and the squared mass parameter \eqref{massivemodemass} provide important features of the massive excitation. For a time-like background vector $b_{\mu}$, the longitudinal mode $\beta$ has a negative mass and represents a tachyonic mode. Considering a space-like vector $b_{\mu}$, albeit the massive mode has a real mass, this excitation is also
a non-physical mode. In fact, assuming $b_{\mu}=(0,0,0,b)$, the constraint \eqref{betaconstrain} implies that the massive mode could propagate as a plane wave with constant amplitude in the $z$ direction. In order to satisfy the asymptotic boundary conditions as $z\rightarrow\pm\infty$, this amplitude must be set to zero. Thus, no configuration can yield to a physical massive propagation \cite{BluhmFung}.


\section{Radiative corrections\label{radiativecorrections}}

In this section, we study the radiative corrections to the two-point vertex function of the $\beta_{\mu}$ field. The main objective is to verify if the mass term associated with the longitudinal mode receives a correction able to modify their non-physical nature.

The classical Lagrangian defined in Eq. \eqref{eq:Lagra1} is not
gauge invariant and to implement its quantization we do not add any
gauge fixing term, and the corresponding Faddeev\textendash{}Popov
ghost fields. Following a perturbative approach, the Lagrangian density
will be written as $\mathcal{L}_{B}=\mathcal{L}_{0}+\mathcal{L}_{int}$,
where
\begin{equation}
\mathcal{L}_{0}=-\frac{1}{4}\beta_{\mu\nu}\beta^{\mu\nu}-\lambda\beta_{\mu}b^{\mu}\beta_{\nu}b^{\nu},\label{L0}
\end{equation}
and
\begin{equation}
\mathcal{L}_{int}=-\frac{\lambda}{4}\left(\beta_{\mu}\beta^{\mu}\beta_{\nu}\beta^{\nu}+4\beta_{\mu}\beta^{\mu}\beta_{\nu}b^{\nu}\right),
\end{equation}
represent the free and interaction terms, respectively. Note that
the possible couplings matter fields are disregarded here.

\begin{figure}[htb]
\begin{eqnarray}
\raisebox{-0.4cm}{\incps{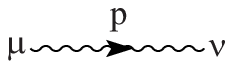}{-1.5cm}{-0.5cm}{1.5cm}{1.5cm}}
 &=& -\frac{i}{p^{2}+i\epsilon}\left[g^{\mu\nu}-\frac{(p^{\mu}b^{\nu}+p^{\nu}b^{\mu})}{b\cdot p}+\frac{(p^{2}+2\lambda b^{2})}{2\lambda(b\cdot p)^{2}}p^{\mu}p^{\nu}\right], \nonumber\\ \\
\raisebox{-0.4cm}{\incps{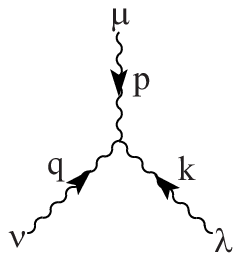}{-1.5cm}{-0.5cm}{1.5cm}{1.5cm}}
 &=& -2i\lambda b^{\alpha}\left(g_{\mu\nu}g_{\alpha\lambda}+g_{\mu\lambda}g_{\nu\alpha}+g_{\mu\alpha}g_{\nu\lambda}\right), \nonumber\\ \\
\raisebox{-0.4cm}{\incps{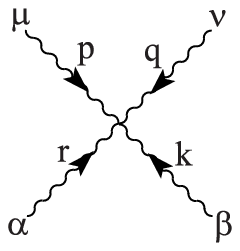}{-1.5cm}{-0.5cm}{1.5cm}{1.5cm}}
 &=& -2i\lambda\left(g_{\mu\nu}g_{\alpha\beta}+g_{\mu\alpha}g_{\nu\beta}+g_{\mu\beta}g_{\nu\alpha}\right). \nonumber\\
\end{eqnarray}
\caption{\label{FeynmanRules}Feynman Rules for the bumblebee model}
\end{figure}

The Feynman rules of this model are summarized in Fig. \eqref{FeynmanRules}.

\begin{figure}[!tb]
\centering
\subfloat[massless tadpole diagram]{
\includegraphics[height=5cm]{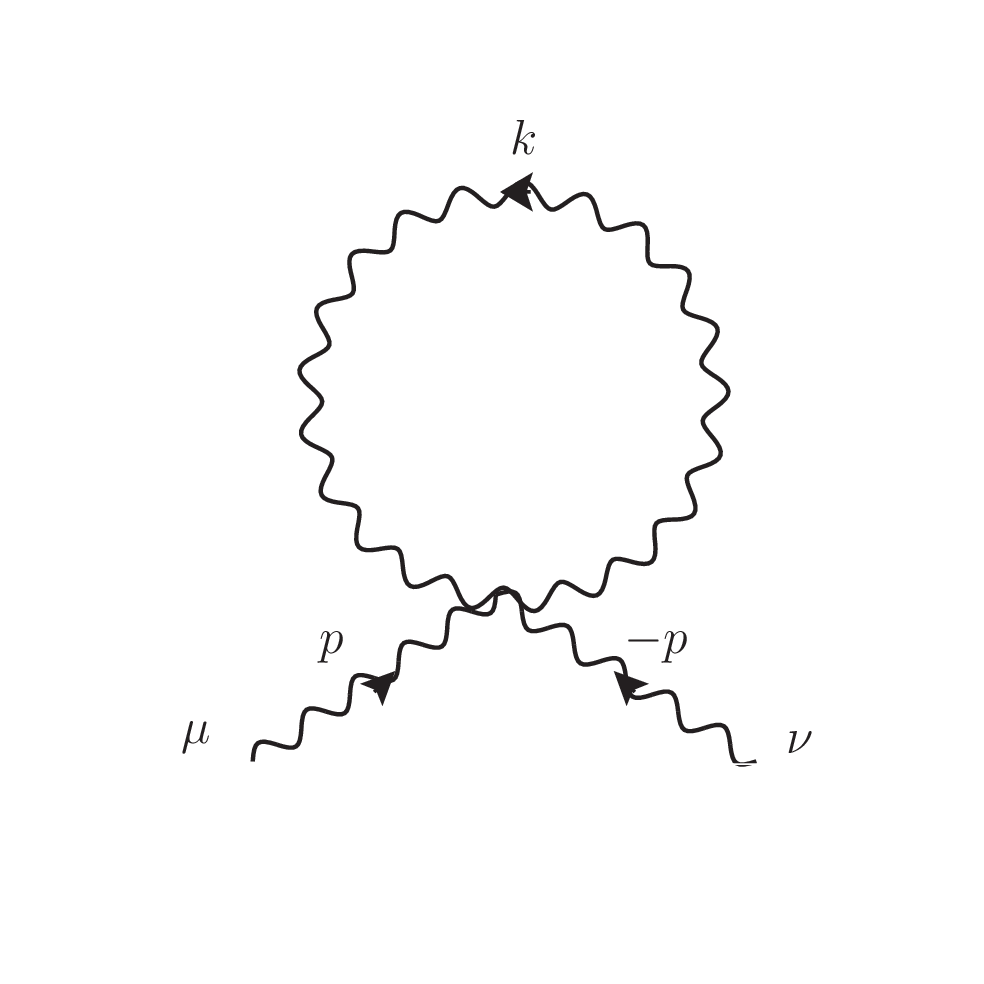}
\label{Diagram1}
}
\subfloat[One-loop self-energy diagram]{
\includegraphics[height=5cm]{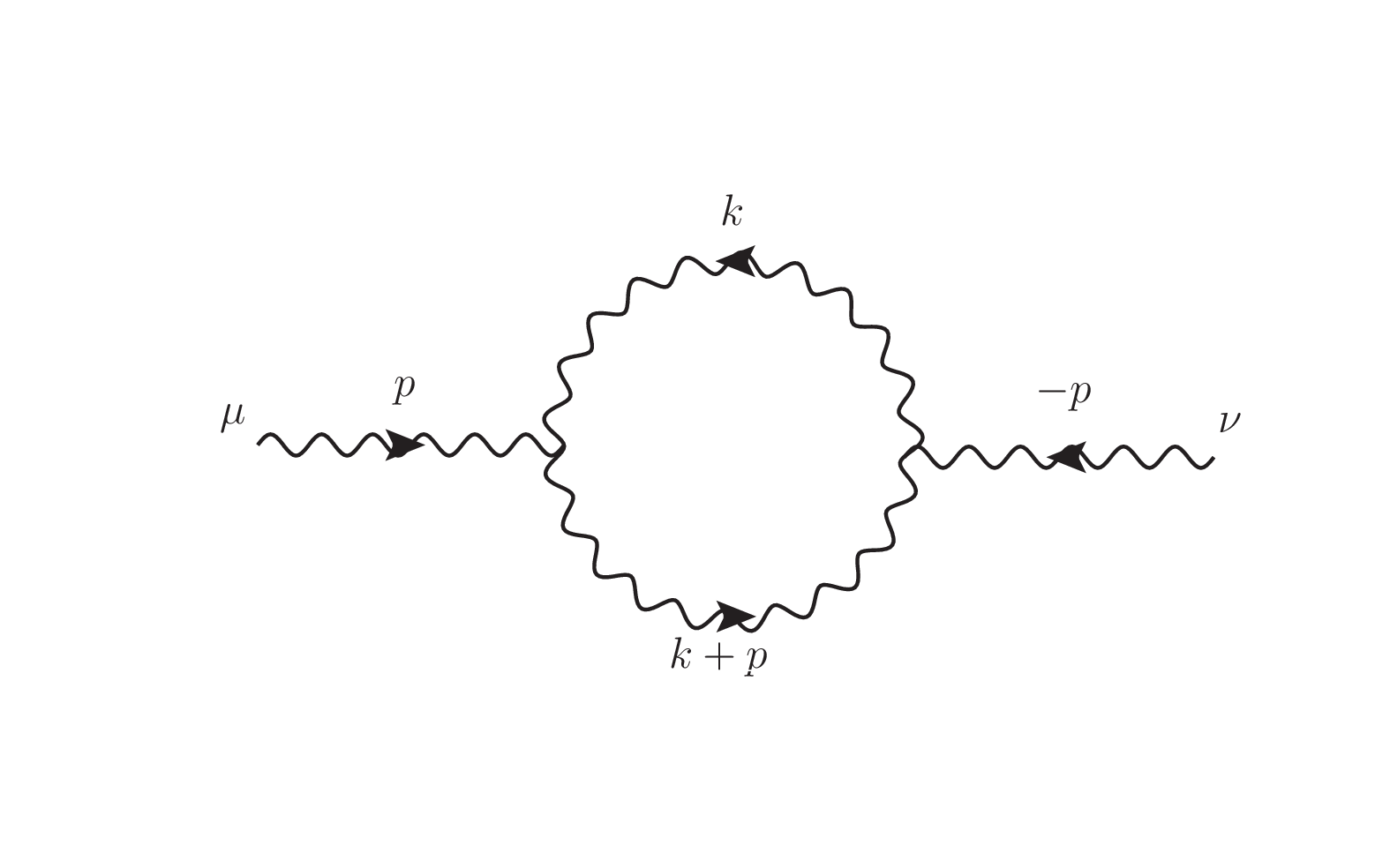}
\label{Diagram2}
}
\caption{Bumblebee two-point vertex functions}
\label{Diagrams}
\end{figure}

As we are interested in the radiative corrections to the bumblebee
mass, we restrict ourselves to the one-loop calculation to the self-energy functions. The relevant diagrams are shown in Fig. \eqref{Diagrams}. These amplitudes are given by
\begin{equation}
\Pi_{\mu\nu}^{(a)}(p,b)=\frac{1}{2}(-2i\lambda)\int\frac{d^{4}k}{(2\pi)^{4}}\left[g_{\mu\nu}D_{\lambda}^{\lambda}(k)+2D_{\mu\nu}(k)\right],
\end{equation}and
\begin{eqnarray}
\Pi_{\mu\nu}^{(b)}(p,b) & = & \frac{1}{2}(-2i\lambda)^{2}b^{\delta}b^{\gamma}\int\frac{d^{4}k}{(2\pi)^{4}}\left[\left(D_{\mu\gamma}(k)D_{\delta\nu}(k+p)+D_{\mu\nu}(k)D_{\delta\gamma}(k+p)\right.\right.\nonumber\\
 & + & \left.D_{\mu\sigma}(k)D_{\delta}^{\sigma}(k+p)g_{\nu\gamma}+D_{\nu\rho}(k)D_{\gamma}^{\rho}(k+p)g_{\mu\delta}+k\leftrightarrow k+p\right)\nonumber\\
 & + & \left.D^{\rho\sigma}(k)D_{\rho\sigma}(k+p)g_{\mu\delta}g_{\nu\gamma}\right].
\end{eqnarray}

It is worthwhile to note that by simple power-counting arguments, the
one-loop integrals above may have ultraviolet divergences up to fourth
order. Furthermore, the presence of unphysical poles $(b\cdot k)^{-\beta}$,
$\beta=1,2...,$ requires a consistent prescription to extract
only the physical content of the theory. A satisfactory prescription for this type of spurious poles has been developed in a long time ago on the quantization of Yang-Mills theory in the axial gauge \cite{pv,pv2,Leibbrandt:1987qv,LeibbrandtBook}. The general axial gauge is defined by the condition $n^{\mu}A^{a}_{\mu}=0$ with $n_{\mu}=(n_{0},{\bf n})$ being a constant four-vector, and it is implemented by the addition of the gauge-fixing Lagrangian density 
\begin{equation*}
\mathcal{L}_{GF}=-\frac{1}{2\alpha}(n\cdot A^{a})^{2},\ \ \alpha\rightarrow0,
\end{equation*} where $\alpha$ is the gauge parameter \cite{Leibbrandt:1987qv}. For this type of gauge, the gauge-field propagator has poles at $n\cdot k=0$ and we can identify the transverse modes \eqref{TransverseMode} with the photon field \cite{Bluhm2005,altschul}. One method that was widely used to deal appropriately with these poles is the so-called principal-value (PV) prescription \cite{LeibbrandtBook}, defined as
\begin{equation}
\frac{1}{(n\cdot k)^{\beta}} \rightarrow\frac{1}{2}\lim_{\mu \rightarrow 0}\left[\frac{1}{(n \cdot k+i\mu)^{\beta}}+\frac{1}{(n\cdot k-i\mu)^{\beta}}\right],
\end{equation} with $\mu>0$ and $\beta=1,2,...,N$. It has been shown that the PV prescription is consistent with the unitarity and renormalization properties of Yang-Mills theories and also preserves the Slavnov-Taylor identities at the one-loop level \cite{pv,pv2}.

For our present purpose, we follow the calculation procedure described in Refs. \cite{Leibbrandt:1987qv,LeibbrandtBook} for the treatment of one-loop axial-type integrals with $n^{2}\neq0$. The corresponding analytical expressions for $\Pi_{\mu\nu}^{(a)}$ and $\Pi_{\mu\nu}^{(b)}$ are shown in \ref{sec:Appendix}. The divergent parts of the momentum integrals are evaluated by dimensional regularization in the PV prescription, whose formulas are presented in Refs. \cite{Leibbrandt:1987qv, LeibbrandtBook}. For simplicity, we consider $\lambda$ small and add all contributions up to order $\lambda$. The diagram (a) turns out to be zero since it is constituted by a tadpole diagram with the massless integrals proportional to
$$\int\frac{d^{D}k}{(k^2)^\alpha(b\cdot k)^\beta},\ \ \alpha,\beta=0,1,2,...$$ that vanishes by dimensional regularization. The non-trivial contribution is only due to the diagram (b), which is given by
\begin{equation}
\Pi_{\mu\nu}(p,b)=\frac{8\lambda}{3}\left(\frac{b\cdot p}{b^{2}}\left(b^{\mu}p^{\nu}+b^{\nu}p^{\mu}\right)-\frac{\left(b^{2}p^{2}+(b\cdot p)^{2}\right)}{b^{4}}b^{\mu}b^{\nu}\right)I_{div}\label{Pi_b},
\end{equation} where $I_{div}=i/8\pi^2\epsilon$ with $\epsilon=4-D$ ($D$ is the dimension of the spacetime). 

We note that unlike the Yang-Mills in the axial gauge, the divergent part of the bumblebee self-energy \eqref{Pi_b} is not transversal, $p_{\mu}\Pi^{\mu\nu}(p,b)\neq0$. The non-transversality of the self-energy can be interpreted as result of the excitation of the massive mode, as verified in the canonical formalism \cite{Hernaski:2014jsa}. On the other hand, no radiative correction to the mass term has been generated for both the NG or the massive mode. Moreover, the one-loop correction has not produced a higher derivative or a non-local divergent term as a function of the external momentum.

Another noteworthy result is that, despite the fact that the bumblebee electrodynamics is superficially nonrenormalizable, no additional counterterm is required to remove the divergences and to make the quadratic effective action finite. Indeed, using the decomposition of the field $\beta_{\mu}$ in transverse and longitudinal modes \eqref{modeTL}, the Lagrangian \eqref{L0} can be rewritten as
\begin{equation}
\mathcal{L}_{0}  = -\frac{1}{4}F^{\mu\nu}F_{\mu\nu}-\frac{1}{2}F^{\mu\nu}\partial_{[\mu}\beta\hat{b}_{\nu]} -\frac{1}{2}\partial_{\mu}\beta\partial^{\mu}\beta+\frac{1}{2}\partial^{\mu}\beta\partial^{\nu}\beta\hat{b}_{\mu}\hat{b}_{\nu}-\lambda b^{2}\beta^{2}.
\end{equation}
In turn, the two-point function $\Pi_{\mu\nu}$ \eqref{Pi_b} contributes to the effective action as
\begin{equation}
\label{effectiveaction}
\mathcal{L}_{div}=\frac{\lambda}{3\pi^{2}}\frac{1}{\epsilon}\left[-\frac{1}{2}F^{\mu\nu}\partial_{[\mu}\beta\hat{b}_{\nu]}-\frac{1}{2}\partial_{\mu}\beta\partial^{\mu}\beta+\frac{1}{2}\partial^{\mu}\beta\partial^{\nu}\beta\hat{b}_{\mu}\hat{b}_{\nu}\right],
\end{equation} such that the divergent terms in \eqref{effectiveaction} can be renormalized by counterterms obtained from the free classical Lagrangian \eqref{L0}.

\section{Conclusions\label{sec:Conclusions}}

In this letter, we addressed the radiative corrections of the bumblebee electrodynamics in flat spacetime. We have chosen the KS model, where the dynamics is governed by a Maxwell-like kinetic term and the Lorentz symmetry is spontaneously broken by a smooth quadratic potential.

The vacuum expected value of the bumblebee field gives rise to a preferred direction, in which the bumblebee excitations can be projected. The transverse mode is massless, and it can be identified with the Nambu-Goldstone mode, whereas the longitudinal mode is massive and represents a non-physical degree of freedom.

Employing the principal value (PV) prescription, which is suitable to deal with the poles of the form $(b\cdot p)$, we obtained the one-loop correction to the bumblebee self-energy. It turned out that the self-energy is not transversal. This is related to the lack of gauge symmetry which allows excitations for the massive mode. Furthermore, the quadratic part of the effective action is free of non-local terms and renormalizable at one-loop order. Even though the smooth quadratic potential breaks the gauge symmetry, no radiative corrections was produced for the NG mode mass or for the massive mode. As a matter of fact, the massive mode remains propagating at one-loop order, which indicates that the non-physical mode detected in the analysis of KS model (tree-level) \cite{BluhmFung,Kostelecky:2008ts} persists at the quantum level.

As future developments of the quantum features of the bumblebee electrodynamics we point out the analysis of process involving coupling with a matter source.
Another noteworthy perspective to be explored is if the requirements to avoid the instabilities in the theory, as carry out in Ref. \cite{Hernaski:2014jsa}, are preserved at the quantum level by loop calculations.

\section*{Acknowledgments}

This work was partially supported by the Brazilian agencies Coordena\c{c}\~ao de Aperfei\c{c}oamento de Pessoal de N\'ivel Superior (CAPES) and Conselho Nacional de Desenvolvimento Cient\'ifico e Tecnol\'ogico (CNPq).  J. E. G. Silva acknowledges the Indiana University Center for Spacetime Symmetries for the kind hospitality.

\appendix
\section{One loop integrals \label{sec:Appendix}}

The analytic expressions for the one-loop self-energy diagrams contributing to the two-point effective action shown in Figs. \eqref{Diagram1} and \eqref{Diagram2} are:

\begin{equation}
\Pi_{\mu\nu}^{(a)}(p,b)=\int\frac{d^{D}k}{(2\pi)^{D}}\left[g^{\mu\nu}\left(-\frac{\lambda b^{2}}{(b\cdot k)^{2}}-\frac{k^{2}}{2(b\cdot k)^{2}}-\frac{\lambda D}{k^{2}}\right)+\frac{2\lambda k^{(\mu}b^{\nu)}}{k^{2}b\cdot k}-\frac{k^{\mu}k^{\nu}\left(2b^{2}\lambda+k^{2}\right)}{k^{2}(b\cdot k)^{2}}\right],\label{diagramA}
\end{equation}and 

\begin{equation}
\Pi_{\mu\nu}^{(b)}(p,b)=\int\frac{d^{D}k}{(2\pi)^{D}}\left[g^{\mu\nu}\Pi_{g}+p^{\mu}p^{\nu}\Pi_{pp}+p^{(\mu}k^{\nu)}\Pi_{pk}+k^{\mu}k^{\nu}\Pi_{kk}+p^{(\mu}b^{\nu)}\Pi_{pb}+k^{(\mu}b^{\nu)}\Pi_{kb}+b^{\mu}b^{\nu}\Pi_{bb}\right],
\label{DiagramB}\end{equation}with the coefficients defined by the following expressions
\begin{align*}
\Pi_{g} & =  \frac{\lambda}{k^{2}}+\frac{\lambda}{(k+p)^{2}},\\
\Pi_{pp} & =  \frac{\lambda b^{2}}{(k+p)^{2}[b\cdot(k+p)]^{2}}+\frac{1}{2[b\cdot(k+p)]^{2}},\\
\Pi_{pk} & = \frac{\lambda b^{2}}{(k+p)^{2}[b\cdot(k+p)]^{2}}+\frac{1}{2b\cdot kb\cdot(k+p)}+\frac{1}{2[b\cdot(k+p)]^{2}},\\
\Pi_{kk} & =  \frac{\lambda b^{2}}{k^{2}(b\cdot k)^{2}}+\frac{\lambda b^{2}}{(k+p)^{2}[b\cdot(k+p)]^{2}}+\frac{1}{b\cdot kb\cdot(k+p)}+\frac{1}{2(b\cdot k)^{2}}+\frac{1}{2[b\cdot(k+p)]^{2}},\\
\Pi_{pb} & =  \frac{\lambda}{k^{2}b\cdot(k+p)}+\frac{\lambda b^{2}k\cdot(k+p)}{b\cdot k(k+p)^{2}[b\cdot(k+p)]^{2}}-\frac{2\lambda}{b\cdot(k+p)(k+p)^{2}}+\frac{k\cdot(k+p)}{2b\cdot k[b\cdot(k+p)]^{2}},\\
\Pi_{kb} & =  -\frac{2\lambda}{k^{2}b\cdot k}+\frac{\lambda}{k^{2}b\cdot(k+p)}+\frac{\lambda b^{2}k\cdot(k+p)}{k^{2}(b\cdot k)^{2}b\cdot(k+p)}+\frac{\lambda b^{2}k\cdot(k+p)}{b\cdot k(k+p)^{2}[b\cdot(k+p)]^{2}}\nonumber \\
 &   +\frac{\lambda}{b\cdot k(k+p)^{2}}-\frac{2\lambda}{(k+p)^{2}b\cdot(k+p)}+\frac{k\cdot(k+p)}{2(b\cdot k)^{2}b\cdot(k+p)}+\frac{k\cdot(k+p)}{2b\cdot k[b\cdot(k+p)]^{2}},\\
\Pi_{bb} & =  \frac{2\lambda^{2}b^{2}}{k^{2}[b\cdot(k+p)]^{2}}+\frac{2\lambda^{2}b{}^{4}[k\cdot(k+p)]^{2}}{k^{2}(k+p)^{2}(b\cdot k)^{2}[b\cdot(k+p)]^{2}}-\frac{4\lambda^{2}b^{2}k\cdot(k+p)}{k^{2}(k+p)^{2}b\cdot kb\cdot(k+p)}\nonumber \\
 &   +\frac{2\lambda^{2}D}{k^{2}(k+p)^{2}}-\frac{4\lambda^{2}}{k^{2}(k+p)^{2}}+\frac{2\lambda^{2}b^{2}}{(b\cdot k)^{2}(k+p)^{2}}+\frac{\lambda b^{2}[k\cdot(k+p)]^{2}}{k^{2}(b\cdot k)^{2}[b\cdot(k+p)]^{2}}\nonumber \\
 &  -\frac{4\lambda k\cdot(k+p)}{k^{2}b\cdot kb\cdot(k+p)}+\frac{\lambda(k+p)^{2}}{k^{2}[b\cdot(k+p)]^{2}}+\frac{\lambda b^{2}[k\cdot(k+p)]^{2}}{(k+p)^{2}(b\cdot k)^{2}[b\cdot(k+p)]^{2}}\nonumber \\
 &   +\frac{\lambda k^{2}}{(b\cdot k)^{2}(k+p)^{2}}-\frac{4\lambda k\cdot(k+p)}{(k+p)^{2}b\cdot kb\cdot(k+p)}+\frac{[k\cdot(k+p)]^{2}}{2(b\cdot k)^{2}[b\cdot(k+p)]^{2}}.
\end{align*}In the above expressions, the round brackets denote symmetrization without the factor $1/2$ (i.e. $a^{(\mu}b^{\nu)}=a^{\mu}b^{\nu}+a^{\nu}b^{\mu}$). 

The divergent parts of the one-loop momentum integrals in \eqref{diagramA} and \eqref{DiagramB} can be performed by using the formulas for axial-type integrals in the PV-prescription from  Appendices in Refs. \cite{Leibbrandt:1987qv, LeibbrandtBook}. The final result up to order $\lambda$ is written in Eq. \eqref{Pi_b}.


\section*{References}
\begin{thebibliography}{99}

\bibitem{noncommutativegeometry}
E. Witten,  Nucl. Phys. B {\bf 268}, 253 (1986); N. Seiberg and E.Witten, J. High Energy Phys. {\bf 9909}, 032 (1999); S. M. Carroll, J. A. Harvey, V. A. Kosteleck\'{y}, C. D. Lane, and T. Okamoto, Phys. Rev. Lett. {\bf 87}, 141601 (2001); Z. Guralnik, R. Jackiw, S. Y. Pi, and A. P. Polychronakos, Phys. Lett. B {\bf 517}, 450 (2001).


\bibitem{horava} P. Ho\v{r}ava, Phys. Rev. D {\bf 79}, 084008 (2009); T. P. Sotiriou, M. Visser and S. Weinfurtner,
J. High Energy Phys. {\bf 0910}, 033 (2009); G. Calcagni, J. High Energy Phys. {\bf 0909}, 112 (2009); M. Visser, Phys. Rev. D {\bf 80}, 025011 (2009).

\bibitem{KS}  V. A. Kosteleck\'{y} and S. Samuel, Phys. Rev. D \textbf{39},
683 (1989); Phys. Rev. Lett. {\bf 63}, 224 (1989); Phys. Rev. D {\bf 40}, 1886 (1989).

\bibitem{espontaneous lorentz violation string} V. A. Kosteleck\'{y} and S. Samuel, Phys. Rev. Lett. {\bf 66}, 1811
(1991); Phys.\ Rev.\ D {\bf 42}, 1289 (1990); V. A. Kosteleck\'{y} and R. Potting, Nucl. Phys. B {\bf 359},
545 (1991); Phys. Lett. B {\bf 381}, 89 (1996); Phys. Rev. D {\bf 63},
046007 (2001); V. A. Kosteleck\'{y}, M. J. Perry, and R.
Potting, Phys. Rev. Lett. {\bf 84}, 4541 (2000).


\bibitem{sme} D. Colladay and V. A. Kosteleck\'{y}, Phys. Rev. D \textbf{55},
6760 (1997); {\bf 58}, 116002 (1998); S. R. Coleman and S. L. Glashow, Phys. Rev. D {\bf 59}, 116008 (1999).


\bibitem{gauge} S. M. Carroll, G. B. Field, and R. Jackiw, Phys. Rev. D {\bf 41},
1231 (1990); R. Jackiw and V. A. Kosteleck\'{y}, Phys. Rev. Lett. {\bf 82}, 3572 (1999); V. A. Kosteleck\'y and M. Mewes, Phys. Rev. Lett. \textbf{87},
251304 (2001); Phys. Rev. D \textbf{66}, 056005 (2002); Phys. Rev. Lett.
\textbf{97}, 140401 (2006); 

\bibitem{gauge2} B. Altschul, Phys. Rev. D {\bf 69}, 125009 (2004); Phys. Rev. D {\bf 73}, 036005 (2006); Phys. Rev. Lett. \textbf{98}, 041603 (2007); 

\bibitem{gauge3} R. Casana, M.M. Ferreira Jr, C. E. H. Santos, Phys. Rev. D
{\bf 78}, 105014 (2008); R. Casana, M.M. Ferreira Jr, A. R. Gomes, P. R. D.
Pinheiro, Eur. Phys. J. C {\bf 62}, 573 (2009); R. Casana, M. M. Ferreira, Jr., A. R. Gomes, and F. E. P. dos Santos, Phys. Rev. D {\bf 82}, 125006 (2010).

\bibitem{gauge4} C. D. Carone, M. Sher, M. Vanderhaeghen, Phys. Rev. D {\bf 74}, 077901 (2006); M. Schreck, Phys. Rev. D \textbf{86}, 065038 (2012).


\bibitem{fermion} W. F. Chen and G. Kunstatter, Phys. Rev. D {\bf 62}, 105029 (2000); G. M. Shore, Nucl. Phys. B{\bf 717}, 86 (2005); D. Colladay and V. A. Kosteleck\'{y}, Phys. Lett. B {\bf 511}, 209 (2001); V. A. Kosteleck\'{y} and R. Lehnert, Phys. Rev. D {\bf 63}, 065008 (2001); R. Lehnert, Phys. Rev. D {\bf 68}, 085003 (2003); 

\bibitem{fermion2} B. Altschul, Phys. Rev. D {\bf 74}, 083003 (2006); B. Goncalves, Y. N. Obukhov, and I. L. Shapiro, Phys. Rev. D {\bf 80}, 125034 (2009). B. Charneski, M. Gomes, R. V. Maluf, and A. J. da Silva, Phys. Rev. D \textbf{86}, 045003 (2012).

\bibitem{KosteleckyG1} V. A. Kosteleck\'{y}, Phys. Rev. D \textbf{69}, 105009
(2004).

\bibitem{Bluhm2005} R. Bluhm and V. A. Kosteleck\'{y}, Phys. Rev. D \textbf{%
71}, 065008 (2005).

\bibitem{BluhmFung} R. Bluhm, S-H. Fung, and V. A. Kosteleck\'{y}, Phys. Rev. D
\textbf{77}, 065020 (2008).


\bibitem{Petrov} A. Ferrari, M. Gomes, J. R. Nascimento, E. Passos, A. Y.
Petrov, and A. J. da Silva, Phys. Lett. B \textbf{652}, 174 (2007).

\bibitem{Pereira2011} B. Pereira-Dias, C. A. Hernaski, and J. A. Helay\"{e}
l-Neto, Phys. Rev. D \textbf{83}, 084011 (2011).

\bibitem{Boldo} J. L. Boldo, J. A. Helay\"{e}l-Neto, L. M. de Moraes, C. A.
G. Sasaki, and V. J. Vasquez Otoya, Phys. Lett. B \textbf{689}, 112 (2010).

\bibitem{MalufGravity1} R. V. Maluf, Victor Santos, W. T. Cruz, and C. A. S.
Almeida, Phys. Rev. D \textbf{88}, 025005 (2013).

\bibitem{MalufGravity2} R. V. Maluf, C. A. S. Almeida, R. Casana, and M. M. Ferreira, Jr., Phys. Rev. D \textbf{90}, 025007 (2014).

\bibitem{PetrovGodel} A.F. Santos, A.Yu. Petrov, W. D. R. Jesus, J. R. Nascimento,  Mod. Phys. Lett. A {\bf 30}, 1550011 (2015).

\bibitem{Kostelecky:2008ts} V. A. Kosteleck\'{y} and N. Russell, Rev. Mod. Phys.  {\bf 83}, 11 (2011).

\bibitem{Seifert} M. D. Seifert, Phys. Rev. D {\bf 81}, 065010 (2010).

\bibitem{altschul} B. Altschul and V. A. Kostelecky, Phys. Lett. B {\bf 628}, 106 (2005).

\bibitem{Bluhm:2008yt} R. Bluhm, N. L. Gagne, R. Potting and A. Vrublevskis, Phys.  Rev. D {\bf 77}, 125007 (2008); Erratum: Phys. Rev. D {\bf 79}, 029902(E) (2009).

\bibitem{Hernaski:2014jsa} C. A. Hernaski, Phys. Rev. D {\bf 90}, 124036 (2014).

\bibitem{Carroll2009} Sean M. Carroll, Timothy R. Dulaney, Moira I. Gresham, and Heywood Tamx, Phys. Rev. D {\bf 79}, 065011 (2009).

\bibitem{pv} W. Kummer, Acta Phys. Austriaca {\bf 41}, 315 (1975). W. Konetschny, and W. Kummer, Nucl. Phys. {\bf 8} 108, 397 (1976); J. Frenkel, and J. C. Taylor, Nucl. Phys. {\bf 8} 109, 439 (1976); 

\bibitem{pv2} D. M. Capper, and G. Leibbrandt, Phys. Rev. D {\bf 25}, 1002 (1982); P. V. Landshoff, Phys. Lett. B {\bf 169}, 69 (1986); P. Gaigg, M. Kreuzer, M. Schweda, and O. Piguet, J. Math. Phys. {\bf 28}, 2781 (1987); P. Gaigg, M. Kreuzer and G. Pollak, Phys. Rev. D {\bf 38}, 2559 (1988).

\bibitem{Leibbrandt:1987qv} G. Leibbrandt, Rev. Mod. Phys.  {\bf 59}, 1067 (1987).
  
\bibitem{LeibbrandtBook} G. Leibbrandt, {\it Noncovariant gauges: Quantization of Yang-Mills and Chern-Simons Theory in Axial-Type Gauges} (World Scientific, Singapore, 1994).





\end{thebibliography}
\end{document}